\documentclass[twocolumn,showpacs,prb,amsmath,amssymb,floatfix]{revtex4-1}

\usepackage{amsmath,amssymb,color}

\usepackage{bm}
\usepackage{epsfig}
\usepackage{psfrag}

\newcommand{\bd}{\bm}

\begin{document}

\title{Summing parquet diagrams using
the functional renormalization group: \\
X-ray problem revisited
}

\author{Philipp Lange, Casper Drukier, Anand Sharma, and Peter Kopietz}
\affiliation{Institut f\"{u}r Theoretische Physik, Universit\"{a}t
  Frankfurt,  Max-von-Laue Strasse 1, 60438 Frankfurt, Germany}

\date{\today}

 \begin{abstract}
We present a simple method for summing  so-called parquet diagrams 
of fermionic many-body systems with competing instabilities  
using  the functional renormalization group.
Our method is based on  partial bosonization of the interaction multi-channel Hubbard-Stratonovich transformations. A simple
truncation of the resulting flow equations, retaining only the frequency-independent parts of the two-point and three-point vertices amounts to
solving coupled Bethe-Salpeter equations
for the effective interaction  to leading logarithmic order.
We apply our method by revisiting the X-ray problem and
deriving the singular frequency dependence of the 
X-ray response function and the particle-particle susceptibility.
Our method is quite general and should be useful in
many-body problems involving strong fluctuations
in several  scattering channels.

\end{abstract}

\pacs{05.10.Cc, 78.70.Dm, 72.15.Qm}

\maketitle

\section{\label{sec:introduction} Introduction}

The basic structure of two-particle Green functions of interacting quantum many-body systems
can exhibit singularities for different combinations of external momenta and frequencies.
In simple cases only one particular combination dominates
the perturbation series, so that it is sufficient to resum the
diagrams corresponding to this combination. 
For example, slightly above a superconducting instability
the two-particle Green function becomes singular if the total momentum
and the total energy of the two incoming particles are small 
(particle-particle channel). It is then sufficient to approximate the effective interaction, appearing in the skeleton diagram of the two-particle Green function, by summing
only the ladder diagrams 
with small total momentum\cite{Fetter71}. 
Another example is the electron gas at high densities, where
the effective interaction 
can be calculated by summing the geometric series
of particle-hole bubbles which dominate for small momentum transfers \cite{Fetter71}.
In some cases, however, a single dominant scattering channel does not exist
and  singularities in more than one scattering channel appear in 
perturbation theory.
In such cases, the usual strategy of summing only particle-particle ladders or 
particle-hole bubbles is inapplicable
and one has to solve coupled Bethe-Salpeter equations
in more than one channel. Diagrammatically, this 
means summing the so-called parquet diagrams where particle-hole bubbles 
and particle-particle ladders
are self-consistently inserted into each other.
Historically,  such a resummation was first used
by Sudakov and co-authors \cite{Sudakov56,Diatlov57} 
in the context of high-energy meson-meson scattering.
Since then,  parquet methods  have played an
important role in studying
quantum impurity models\cite{Abrikosov65,Roulet69,Nozieres69,Fukushima71,kleinert95,Janis99},
low-dimensional metals \cite{Dzyaloshinskii72,Gorkov74,Dzyaloshinskii87,Zheleznyak97},
liquid Helium \cite{Babu73,Jackson82,Quader87,Pfitzner87}, and vortex liquids\cite{Yeo96}.
Moreover parquet methods have also been used to construct approximations to reduced density matrices of large physical systems\cite{Yasuda99}, 
nuclear structure calculations\cite{Bergli10},
and different lattice models for strongly correlated 
electrons\cite{Bickers91,Bickers92,Irkhin01,Chubukov08,Maiti10,Maiti13,Nandkishore12,Tam13}.
In particular, in two seminal papers by
Roulet {\it{et al.}} \cite{Roulet69}
and Nozi\`{e}res {\it{et al.}} \cite{Nozieres69}
the threshold exponents of the so-called X-ray problem\cite{Mahan67} 
were obtained using the parquet method. The agreement of the parquet result for the
threshold exponents  with the known exact result \cite{Nozieres69b} at weak coupling 
supports the validity of this technique.

While parquet methods are a well established many-body tool\cite{Smith88,Gogolin99},
they are not straightforward to apply to physical problems of interest.
One reason could be related to the fact
that the derivation of coupled Bethe-Salpeter equations
in several channels often relies on subtle diagrammatic
and combinatoric considerations.
Moreover, the explicit solution of coupled Bethe-Salpeter equations 
often requires rather substantial algebraic manipulations or extensive numerical calculations. 
It is well known, however, that 
the summation of parquet diagrams can also be formulated
in terms of the renormalization group.
In fact,  the modern formulation of the
Wilsonian renormalization group for fermionic many-body systems 
in terms of a formally exact
hierarchy of flow equations for the one-particle irreducible 
vertices \cite{Kopietz01,Salmhofer01,Kopietz10,Metzner12} 
(we shall refer to this approach as the
functional renormalization group, abbreviated by FRG) 
opens new possibilities for dealing with parquet type of problems.
On the one hand, the FRG offers a general and systematic framework of
deriving parquet equations in a purely algebraic way without 
relying on diagrammatic arguments; on the other hand, the formulation
of  coupled Bethe-Salpeter equations as FRG
flow equations offers new strategies of obtaining approximate solutions.

In this work we shall revisit the X-ray problem and present a systematic approach to obtain the well-known parquet results \cite{Roulet69,Nozieres69}
within the framework of the FRG. In the X-ray problem both the particle-particle and the particle-hole channel exhibit logarithmic singularities, so 
that the calculation of the X-ray response function can be used as a benchmark for testing truncations of the FRG flow equations. Moreover the approximations made in 
Refs.~[\onlinecite{Roulet69,Nozieres69}] rely 
on the logarithmic nature of the singularities. 
However, the necessary algebra is quite demanding.
Here we show that the threshold exponents of the
X-ray problem and the frequency dependence of the corresponding response function
can be obtained in a relatively straightforward manner within the FRG. 
Our particular implementation of the FRG relies on the partial bosonization
of the theory, in both singular scattering channels, using multi-channel Hubbard Stratonovich transformations. 
We believe that this technique will also be useful in other problems which are
dominated by competing singularities in several channels.

The rest of the paper is organized as follows. In section~\ref{sec:xrayproblem} 
we briefly present the X-ray problem and formulate it using functional integrals. 
In section~\ref{sec:deephole} we identify the cut-off scheme for the FRG and introduce the regularized 
deep hole Green function. We write down the exact FRG flow equation for the deep hole self-energy and 
calculate the leading term in the weak coupling expansion of the anomalous dimension of the {\it{d}}-electrons. 
In the succeeding section~\ref{sec:responsefunction}, we present our FRG-based approach to obtain the particle-hole susceptibility, 
describing the X-ray response, and also derive a similar expression for the corresponding particle-particle susceptibility. 
In the concluding section~\ref{sec:conclusions} we summarize our obtained results.

\section{\label{sec:xrayproblem} Functional integral formulation of the X-ray problem}

Before we present the X-ray problem in the functional integral framework, we shall briefly describe the nature of the problem. The X-ray absorption and emission in metals
is modeled by the following second quantized Hamiltonian \cite{Mahan67}
 \begin{equation}
 {\cal{H}} = \sum_{\bd{k}} \epsilon_{\bd{k}} c^{\dagger}_{\bd{k}} c_{\bd{k}}
 + \epsilon_d d^{\dagger} d + \sum_{\bd{k} \bd{k}^{\prime}} U_{\bd{k} \bd{k}^{\prime}} 
 c^{\dagger}_{\bd{k}} c_{\bd{k}^{\prime}} d^{\dagger} d ,
 \label{eq:Hamilton}
 \end{equation}
where $c^{\dagger}_{\bd{k}}$ creates a conduction electron 
with momentum $\bd{k}$ and energy $\epsilon_{\bd{k}} = \bd{k}^2/(2m )$, while
$d^{\dagger}$ creates a localized deep core electron with energy $\epsilon_d < 0$.
The Coulomb interaction between the conduction electrons is neglected and only 
the intraband part of the Coulomb
interaction between the conduction electrons and the deep core electron is taken into account.
For simplicity, the spin degrees of freedom are ignored and 
a separable interaction of the form
$U_{\bd{k} \bd{k}^{\prime}} = U u_{\bd{k}} u_{\bd{k}^{\prime}}$
is assumed, where the form factors $u_{\bd{k}}$ will be specified below.

We are interested in the 
$d$-electron Green function 
 \begin{equation}
 G^d ( t - t^{\prime} ) = - i \langle {\cal{T}} [ d (t ) d^{\dagger} ( t^{\prime} ) ]
 \rangle.
 \end{equation}
Moreover, the
experimentally measurable X-ray transition rate 
is determined by the  Fourier transform of the particle-hole response function
 \begin{equation}
 \chi^{\rm ph}  ( t - t^{\prime} ) = \langle {\cal{T}} [ \hat{A} ( t ) 
\hat{A} (t^{\prime} ) ] \rangle,
 \end{equation}
where ${\cal{T}}$ is the time ordering symbol and
the composite particle-hole operator 
$\hat{A} ( t )$ is defined by
\begin{equation}
\hat{A} ( t ) =  [ c^{\dagger} ( t ) d ( t )
 + d^{\dagger} ( t ) c ( t ) ] ,
 \end{equation}
with
 \begin{equation}
c ( t ) = \sum_{\bd{k}} u_{\bd{k}} c_{\bd{k}} ( t ).
 \label{eq:cdef}
 \end{equation}
The Fourier transforms $G^d ( \omega )$ and
 $\chi^{\rm ph} ( \omega )$ of the above functions
for real frequencies $\omega $ in the vicinity of
certain threshold frequencies $\omega_d$ and $\omega_x$ are known to be of the form \cite{Mahan67,Roulet69,Nozieres69}
 \begin{eqnarray}
 G^d ( \omega ) & \sim  & \frac{1}{ \omega - \omega_d } 
 \left( \frac{ \omega - \omega_d }{\xi_0} \right)^{ \eta},
 \label{eq:Gdthresh}
 \\
 \chi^{\rm ph} ( \omega ) & \sim & \frac{\nu_0}{2 u} 
 \left[  \left( \frac{  \xi_0}{\omega - \omega_x } \right)^{\alpha} -
 1 \right],
 \label{eq:responsealpha}
 \end{eqnarray}
where $\nu_0$ is the density of states
of conduction electrons at the Fermi energy and $\xi_0$ is an ultraviolet cutoff
of the order of the width of the conduction band.
For small values of the  dimensionless interaction $u = \nu_0 U$
the threshold exponents are given by
 \begin{eqnarray}
 \eta & = & u^2 + {\cal{O}} ( u^3),
 \label{eq:weaketa}
 \\
 \alpha & = &  2 u + {\cal{O}} ( u^2).
 \end{eqnarray}
In Refs.~[\onlinecite{Roulet69,Nozieres69}] the above results were derived by explicitly writing down and solving parquet equations 
which resum the leading logarithmic singularities in both the
particle-hole and the particle-particle channel.
In the following we show that the FRG offers an alternative and, in our opinion, simpler
way to derive these results.

In order to set up the FRG for the X-ray problem, it is useful to start from an effective
Euclidean action depending only on the degrees of freedom corresponding
to the $d$-electrons and the relevant linear combination (\ref{eq:cdef})
of the conduction electron operators.
To derive this, consider the Euclidean action
associated with the Hamiltonian (\ref{eq:Hamilton}),
 \begin{eqnarray}
 S  & =   &  \int_{0}^{\beta} d \tau  \sum_{\bd{k} }   \bar{c}_{\bd{k}} ( \tau ) 
( \partial_{\tau} + \xi_{\bd{k}} ) c_{\bd{k}} ( \tau )
 \nonumber
 \\
 & + & \int_0^{\beta} d \tau \bar{d} (\tau ) ( \partial_{\tau} + \xi_d ) d ( \tau )
 \nonumber
 \\
 & + & U \int_{0}^{\beta} d \tau \bar{c} ( \tau ) c ( \tau ) \bar{d} ( \tau ) d ( \tau ),
 \label{eq:Sdef}
\end{eqnarray} 
where the energies 
$\xi_{\bd{k}} = \epsilon_{\bd{k}} - \mu$ and  $\xi_d = \epsilon_d - \mu$ are measured
relative to the chemical potential $\mu$. 
Although at this point  we keep the 
inverse temperature $\beta$ finite, 
later we will take the zero temperature limit $\beta \rightarrow \infty$ whenever it is convenient. 
The quantities $c_{\bd{k}} (\tau )$ and $d ( \tau )$ are now Grassmann variables 
depending on imaginary time $\tau$, 
and the Grassman variable $c (\tau)$ represents the linear combination
$ c ( \tau ) = \sum_{\bd{k}} u_{\bd{k}} c_{\bd{k}} ( \tau )$.
To derive an effective action depending only on
$c ( \tau )$, we integrate over all $c_{\bd{k}}$-variables with the exception of the
linear combination $c ( \tau )$ of interest. 
Thus we
insert the following representation of unity 
into the functional integral representation of the correlation functions,
 \begin{eqnarray}
 1 & = & \int {\cal{D}} [ \bar{c}^{\prime} , c^{\prime} ]
 \prod_{\tau} 
 \delta ( \bar{c}^{\prime} ( \tau )   - \bar{c} (\tau ) ) 
 \delta ({c}^{\prime} ( \tau )   - {c} (\tau ) ) 
 \nonumber
 \\
 & = &  \int {\cal{D}} [ \bar{c}^{\prime} , c^{\prime} ] 
\int {\cal{D}} [ \bar{\eta} , \eta ]  e^{ \int_0^{\beta} d \tau [
 \bar{\eta} ( c^{\prime} - c ) + ( \bar{c}^{\prime} - \bar{c} ) \eta ]},
 \label{eq:deltagras}
 \end{eqnarray}
where $c^{\prime}$ and $\eta$ are new Grassmann variables and
the functional delta-function of the Grassmann variables 
is defined as usual \cite{Shifman99}.
Then we integrate over the $c_{\bd{k}}$-variables and subsequently
over the auxiliary $\eta$-variables. After renaming 
$c^{\prime} \rightarrow c $ we Fourier transform the Gaussian part 
to frequency space and obtain the effective impurity model
 \begin{eqnarray}
 S_{\rm imp} & = & - \int_{\omega} (G_{0}^{c}( i \omega ))^{-1} \bar{c}_{\omega} c_{\omega}
   - \int_{\omega} (G_{0}^{d}( i \omega ))^{-1} \bar{d}_{\omega} d_{\omega}
  \nonumber
 \\
 & + & U \int_0^{\beta} d \tau \bar{c} ( \tau ) c ( \tau ) \bar{d} ( \tau ) d ( \tau ),
 \label{eq:Simp}
 \end{eqnarray}
where the symbol $ \int_{\omega}= \beta^{-1} \sum_{\omega}$ denotes summation over fermionic
Matsubara frequencies $i \omega$ such that for vanishing temperature $\int_{\omega} = \int \frac{ d \omega}{2 \pi}$, and
the Fourier components of the fields are defined by
 $c_{\omega} = \int_0^{\beta} d \tau  e^{ i \omega \tau } c ( \tau )$ and
$d_{\omega} = \int_0^{\beta} d \tau  e^{ i \omega \tau } d ( \tau )$. The
Gaussian part of the action (\ref{eq:Simp}) depends on the  
non-interacting Green functions
 \begin{eqnarray}
 G_{0}^c ( i \omega ) & = & \sum_{\bd{k}} \frac{ u_{\bd{k}}^2}{ i \omega - \xi_{\bd{k}} },
  \\
 G_{0}^d ( i \omega ) & = & \frac{1}{ i \omega - \xi_d }.
 \end{eqnarray}
At this point  it is convenient to assume that the form factor
$u_{\bd{k}}$  is  only finite for
$|\xi_{\bd{k}} | < \xi_0$,  where $\xi_0$ is a bandwidth  cutoff
of the order of the Fermi energy $\epsilon_F = \mu$.
Moreover, we also assume that $u_{\bd{k}}$ is such that 
 \begin{equation}
 \sum_{\bd{k}} \frac{ u_{\bd{k}}^2}{ i \omega - \xi_{\bd{k}} } = \nu_0
 \int_{ - \xi_0}^{\xi_0} d \xi \frac{1}{i \omega - \xi },
 \end{equation}
where $\nu_0$ is the density of states at the Fermi energy.
The above integral becomes elementary and we obtain
 \begin{equation}
 G_{0}^c ( i \omega ) = - i \pi \nu_0 {\rm sgn} \omega 
 \left[ 1 - \frac{2}{\pi} \arctan \left( \frac{| \omega | }{ \xi_0} \right) \right].
 \label{eq:Gcres}
 \end{equation}
The effective impurity model defined by
Eq.~(\ref{eq:Simp}) is the starting point of our FRG calculation.

\section{\label{sec:deephole} Deep hole Green function}

\begin{figure}[t]
\includegraphics[width=80mm]{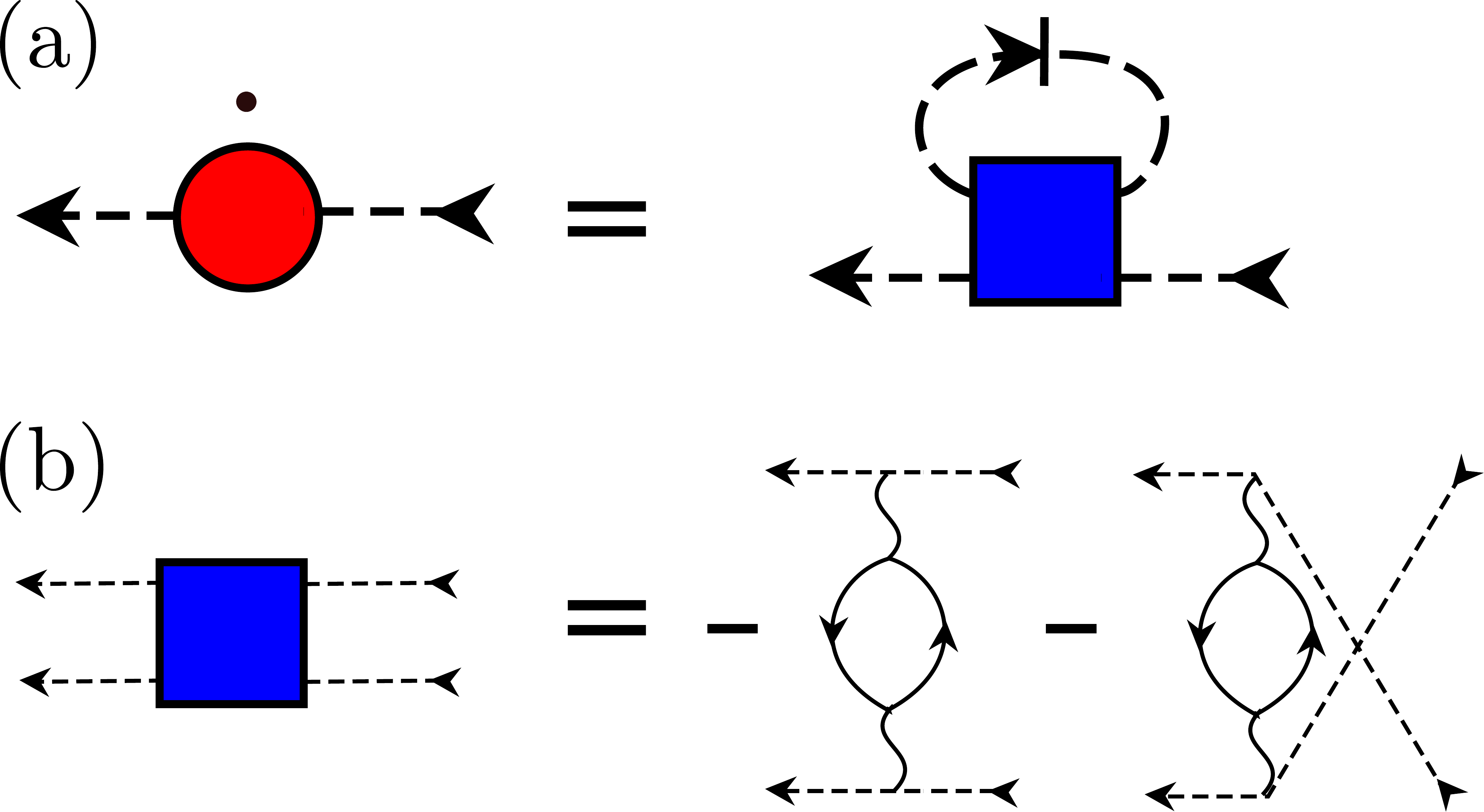}

  \caption{%
(Color online)
(a)
Graphical representation of the exact FRG flow equation (\ref{eq:FRG})
for the $d$-electron self-energy. The dashed arrows denote the $d$-electron propagators,
the dashed arrow with an extra slash denotes the $d$-electron single-scale propagator,
the shaded (red) circle with a dot represents 
the scale derivative of the $d$-electron self-energy,
and the shaded (blue) square represents the  one-particle irreducible effective
interaction between the $d$-electrons.
(b) Feynman diagrams contributing to the effective interaction
(\ref{eq:Gamma4}) between the $d$-electrons to second order in the bare interaction.
The solid arrows forming the loops 
represent the non-interacting conduction electron propagators.
}
\label{fig:flowd}
\end{figure}

We begin by discussing the FRG calculation of the $d$-electron (deep hole) 
Green function. It turns out that the leading order term in the weak coupling expansion of
the corresponding anomalous dimension $\eta$ can be obtained 
in a straightforward way
without resumming parquet diagrams.
To set up the FRG procedure, we 
need to specify our cutoff scheme.
Since the propagator of the $c$-fermions does not exhibit any singularity [see Eq.~\eqref{eq:Gcres}],
we do not introduce any cutoff in this sector.
On the other hand, we regularize the singularity of the
non-interacting $d$-electron propagator by replacing
 \begin{equation}
 ( G_0^d ( i \omega ) )^{-1} \rightarrow 
 ( G_{0 \Lambda} ^d ( i \omega ) )^{-1} = i \omega - \xi_d + R^d_{\Lambda} ( i \omega ),
 \end{equation}
where the regulator function 
$R^d_{\Lambda} ( i \omega )$
vanishes for $\Lambda \rightarrow 0$ and
diverges for $\Lambda \rightarrow \infty$.
The regularized $d$-electron propagator  is then
 \begin{equation}
 G^d_{\Lambda} ( i \omega ) = \frac{1}{ i \omega - \xi_d - \Sigma_{\Lambda}^d ( i \omega )
 + R^d_{\Lambda} ( i \omega )},
 \end{equation}
where the cutoff-dependent $d$-electron
self-energy  $\Sigma_{\Lambda}^d ( i \omega )$
satisfies the exact FRG flow equation \cite{Kopietz01,Salmhofer01,Kopietz10,Metzner12}
 \begin{eqnarray}
 \partial_{\Lambda} \Sigma^d_{\Lambda} ( i \omega ) & = & \int_{\omega^{\prime}} \dot{G}^d_{ \Lambda}
 ( i \omega^{\prime} )  
 \Gamma_{\Lambda}^{\bar{d} \bar{d} dd} ( 
 \omega , \omega^{\prime} ; \omega^{\prime} , \omega ).
 \label{eq:FRG}
 \end{eqnarray}
Here
\begin{equation}
 \dot{G}^d_{\Lambda} ( i \omega )  = [ - \partial_{\Lambda} R^d_{\Lambda} ( i \omega ) ]
 [ G^d_{\Lambda} ( i \omega ) ]^2
 \end{equation}
is the so-called single-scale propagator.
The irreducible four-point vertex
 $\Gamma_{\Lambda}^{\bar{d} \bar{d} dd} ( 
 \omega_1^{\prime} , \omega_2^{\prime} ; \omega_2 , \omega_1 )$
can be identified with the properly antisymmetrized  effective interaction 
between two $d$-electrons for a given value of the cutoff $\Lambda$.
A graphical representation of the flow equation (\ref{eq:FRG}) is shown in 
Fig.~\ref{fig:flowd} (a).
A convenient regulator for the $d$-electrons is a Litim regulator \cite{Litim01}
in frequency space,
 \begin{equation}
 R^d_{\Lambda} ( i \omega )  =    i  {\rm sgn} \omega
 ( \Lambda - | \omega | )   
 \Theta (  \Lambda - | \omega | ),
 \label{eq:regulator}
 \end{equation}
so that the cutoff derivative of the regulator is 
 \begin{equation}
 \partial_{\Lambda} R^d_{\Lambda} ( i \omega )  =
  i {\rm sgn} \omega \Theta ( \Lambda - | \omega | ).
 \end{equation}
For simplicity we choose the zero of energy such that the
$d$-electron energy vanishes. Ignoring self-energy corrections
to the $d$-electron propagator, we get
  \begin{eqnarray}
 G^d_{\Lambda} ( i \omega )
 & \approx &   \left\{ \begin{array}{cc}
\frac{1}{i \Lambda {\rm sgn} \omega 
 }  & \mbox{for $ | \omega | < \Lambda $},
 \\
 \frac{1}{i \omega 
 }  &  \mbox{for $ | \omega | > \Lambda$} ,
 \end{array}
 \right.
 \end{eqnarray}
and
 \begin{eqnarray}
 \dot{G}^d_{\Lambda} ( i \omega ) \approx \frac{i {\rm sgn} \omega}{\Lambda^2} \Theta (  \Lambda - 
 | \omega | ).
 \label{eq:Gdotd}
 \end{eqnarray}
Since we do not impose any cutoff on the $c$-fermion propagator,
the effective interaction 
 $\Gamma_{\Lambda}^{\bar{d} \bar{d} dd} ( 
 \omega_1^{\prime} , \omega_2^{\prime} ; \omega_2 , \omega_1 )$
between the $d$-electrons
is finite even at the initial scale $\Lambda = \Lambda_0$. It should be pointed out that
our bare action (\ref{eq:Sdef}) does not contain an interaction of this type.
Diagrammatically, the effective interaction 
 $\Gamma_{0}^{\bar{d} \bar{d} dd} ( 
 \omega_1^{\prime} , \omega_2^{\prime} ; \omega_2 , \omega_1 )$
at the inital scale $\Lambda= \Lambda_0$ is determined by  all diagrams 
with four external $d$-legs and only
$c$-propagators in internal loops.
In the weak coupling regime,
we may approximate the flowing
 $\Gamma_{\Lambda}^{\bar{d} \bar{d} dd} ( 
 \omega_1^{\prime} , \omega_2^{\prime} ; \omega_2 , \omega_1 )$,
on the right-hand side of Eq.~(\ref{eq:FRG}), by its
 initial value at $\Lambda_0$.
For the calculation of the leading term in the weak coupling expansion of
the anomalous dimension $\eta$ of the $d$-electrons,
we only need the effective interaction to second order in the bare interaction $U$.
For the special frequencies
$\omega_1^{\prime} = \omega_1 = \omega$ and 
$\omega_2^{\prime} = \omega_2 = \omega^{\prime}$
appearing on the right-hand side of the FRG flow equation (\ref{eq:FRG})
we obtain the effective interaction between the d-electrons up to second order in the bare interaction
 \begin{equation}
 \Gamma_{0}^{\bar{d} \bar{d} dd} ( 
 \omega , \omega^{\prime} ; \omega^{\prime} , \omega ) = - U^2 [
 \Pi_0^{cc} ( i \omega - i \omega^{\prime} ) - \Pi_0^{cc} (0) ] + {\cal{O}} ( U^3 ),
 \label{eq:Gamma4}
 \end{equation}
where  $\Pi_0^{cc} ( i \bar{\omega}  )$ is the particle-hole bubble
with non-interacting $c$-fermion propagators,
 \begin{equation}
  \Pi_0^{cc} ( i \bar{\omega}  ) = \int_{\omega} G^c_0 ( i \omega - i \bar{\omega} )
 G^c_0 ( i \omega ).
 \end{equation}
The Feynman diagrams contributing to Eq.~(\ref{eq:Gamma4}) are shown in 
Fig.~\ref{fig:flowd} (b).
For small $| {\omega} |$ we may approximate 
$G_0^{c} ( i \omega ) \approx - i \pi \nu_0 {\rm sgn} \omega$ and obtain
 \begin{equation}
  \Gamma_{0}^{\bar{d} \bar{d} dd} ( 
 \omega , \omega^{\prime} ; \omega^{\prime} , \omega ) = - \pi ( \nu_0 U )^2 |
 \omega - \omega^{\prime} | + {\cal{O}} ( U^3 ).
  \label{eq:Gamma4pert}
 \end{equation}
Substituting this effective interaction into the right-hand side of
the exact FRG flow equation (\ref{eq:FRG}) we obtain a closed integro-differential equation 
for the frequency dependent $d$-electron self-energy $\Sigma^d_{\Lambda} (  i \omega )$. 
From the numerical solution of this equation one can obtain the 
$d$-electron propagator $G^{d}_{\Lambda} ( i \omega )$ for all frequencies.
Fortunately, the singular behavior close to the threshold can be
extracted from the leading term in the low-energy expansion of the self-energy, 
 \begin{equation}
 \Sigma^d_{\Lambda} ( i \omega ) = \Sigma^d_{\Lambda} ( 0 ) - ( 1 - Z_{\Lambda}^{-1} )
 i \omega + {\cal{O}} ( \omega^2 ),
 \label{eq:flowself}
 \end{equation}
where the cutoff-dependence of the wave-function renormalization factor $Z_{\Lambda}$ 
defines the flowing anomalous dimension,
\begin{equation}
 \eta_{\Lambda} =  \Lambda \partial_{\Lambda} \ln Z_{\Lambda}.
 \end{equation}
This quantity
can be obtained from the right-hand side of the exact FRG flow equation 
as follows \cite{Kopietz10},
 \begin{eqnarray}
 \eta_{\Lambda} & = &  Z_{\Lambda} \Lambda \lim_{\omega \rightarrow 0}
 \frac{ \partial}{\partial (i \omega ) } \partial_{\Lambda} \Sigma^d_{\Lambda} ( i \omega )
 \nonumber
 \\
 & = & 
 Z_{\Lambda} \Lambda
\int_{\omega^{\prime}} \dot{G}^d_{ \Lambda}
 ( i \omega^{\prime} )  
 \lim_{\omega \rightarrow 0}
 \frac{ \partial}{\partial (i \omega ) }
 \Gamma_{\Lambda}^{\bar{d} \bar{d} dd} ( 
 \omega , \omega^{\prime} ; \omega^{\prime} , \omega ).
 \label{eq:etaexact}
 \hspace{7mm}
 \end{eqnarray}
Substituting our perturbative second order result (\ref{eq:Gamma4pert})
and our lowest order approximation (\ref{eq:Gdotd}) for the single-scale
propagator in Eq.~(\ref{eq:etaexact}) we obtain 
 \begin{equation}
 \eta = \lim_{\Lambda \rightarrow 0} \eta_{\Lambda} 
 = ( \nu_0 U )^2 + {\cal{O}} ( U^3 ),
 \end{equation}
in agreement with the known weak coupling expansion  (\ref{eq:weaketa}).
Finally, keeping in mind that our low-energy expansion (\ref{eq:flowself})
is only valid for $ | \omega | \lesssim \Lambda$,
we may estimate the  $d$-electron propagator from 
 \begin{equation}
 G^d ( i \omega )  \approx \frac{ Z_{\Lambda = | \omega | }}{ i \omega }
  \approx  \frac{ ( | \omega | / \Lambda_0 )^{\eta}}{ i \omega } .
 \label{eq:Gdlow}
\end{equation}
Recalling that we have set the
threshold energy $\omega_d$ equal to zero, 
we recover
the known threshold behavior (\ref{eq:Gdthresh}) of the $d$-electron propagator,
where we identify $\xi_0 = \Lambda_0$.

\section{\label{sec:responsefunction} X-ray and particle-particle response}

In this section, using the FRG formalism, we demonstrate how to obtain the non-perturbative 
result (\ref{eq:responsealpha}) for the particle-hole susceptibility
$\chi^{\rm ph} ( \omega )$ 
which describes the X-ray response. Moreover, we shall also derive an analogous 
expression for the corresponding particle-particle
susceptibility $\chi^{\rm pp} ( \omega )$.

\begin{figure}[t]
\includegraphics[width=80mm]{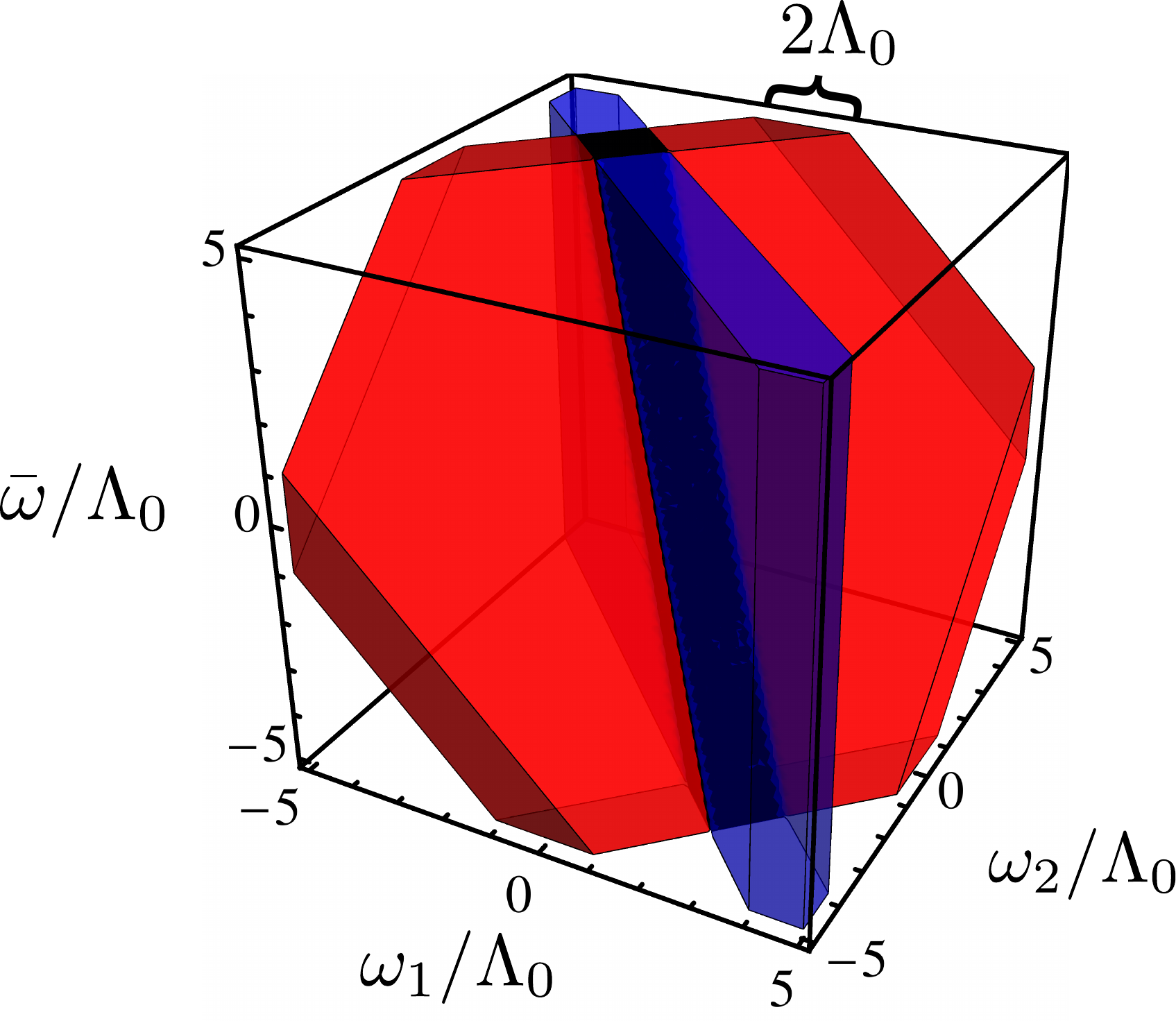}
  \caption{%
(Color online)
Low-energy sectors in the
three dimensional frequency space spanned by
the two frequencies $ \omega_1$ and $\omega_2$ of the incoming fermions and
the frequency transfer $\bar{\omega} = \omega_1^{\prime} - \omega_1$.
The blue sector perpendicular to the $\omega_1 - \omega_2$-plane
represents the regime 
$ | \omega_1 + \omega_2 | < \Lambda_0$
responsible for the dominant renormalization of 
effective interaction in the Cooper channel,
while the red sector 
defined by
$ | \omega_2^{\prime} - \omega_1 | < \Lambda_0$ 
is responsible for the dominant renormalization in the exchange scattering channel.
}
\label{fig:slices}
\end{figure}

To begin with, let us consider the non-interacting
particle-hole and particle-particle bubbles involving
one $c$-fermion and one $d$-electron propagator,
 \begin{eqnarray}
 \Pi_0^{\rm ph} ( i \bar{\omega} ) & = & \int_{\omega} G_{0}^c ( i \omega )
 G_{0}^d ( i \omega - i \bar{\omega} ),
 \label{eq:bubbleph}
 \\
 \Pi_0^{\rm pp} ( i \bar{\omega} ) & = & \int_{\omega} G_{0}^c ( i \omega )
 G_{0}^d ( - i \omega + i \bar{\omega} ).
 \label{eq:bubblepp}
 \end{eqnarray}
Both bubbles exhibit logarithmic singularities. 
To extract these, we note that it is sufficient to retain
only the low-energy part of the
local c-fermion Green function, 
Eq.~(\ref{eq:Gcres}) 
  \begin{equation}
 G_{0}^c ( i \omega ) \approx - i \pi \nu_0 {\rm sgn} \omega \Theta ( \xi_0 - | \omega | ).
 \end{equation}
At vanishing temperature the integrations in Eqs.~(\ref{eq:bubbleph}) and 
(\ref{eq:bubblepp}) can then be performed exactly and we obtain
 \begin{eqnarray}
 \Pi_0^{\rm ph} ( i \bar{\omega} ) & = & - \nu_0 \ln \left[ \frac{ \xi_0}{i \bar{\omega} +
 \xi_d} \right],
 \\
  \Pi_0^{\rm pp} ( i \bar{\omega} ) & = &  \nu_0 \ln \left[ \frac{ \xi_0}{i \bar{\omega} -
 \xi_d} \right].
 \end{eqnarray}
Due to the divergence in both channels, a simple one-channel trunction
is not sufficient to calculate
the X-ray response. The parquet method is a systematic 
tool to resum the leading divergencies in both channels.
There are different implementations of this method. 
Our FRG formulation is closely related to the 
particular implementation of the parquet renormalization group
discussed by Maiti and Chubukov \cite{Maiti13}, who represent
all potentially important fluctuations via bosonic
Hubbard-Stratonovich fields, calculate the  renormalization of the
associated three-legged vertices, and finally use these vertices to
calculate the susceptibilities.
In order to formulate this program within the framework of the FRG,
we bosonize the interactions in the two relevant channels.
To do this, it is sufficient to retain only those interaction processes which 
mediate the singular interactions
in the particle-hole and particle-particle channels.
To derive the corresponding channel decomposition, 
we note that in frequency space the interaction 
in Eq.~(\ref{eq:Sdef})
can be written in the following three equivalent ways,
 \begin{subequations}
 \begin{eqnarray}
 S_{\rm int} & \equiv & 
 U \int_0^{\beta} d \tau \bar{c} ( \tau ) c ( \tau ) \bar{d} ( \tau ) d ( \tau )
 \nonumber
 \\
 & = & U \int_{\omega_1^{\prime}}  \int_{\omega_2^{\prime}}  \int_{\omega_2} \int_{\omega_1}
 \delta_{ \omega_1^{\prime} + \omega_2^{\prime} , \omega_2 + \omega_1 }
 \bar{c}_{\omega_1^{\prime}} \bar{d}_{\omega_2^{\prime}} d_{\omega_2} c_{\omega_1}
\nonumber
 \\
 & = & 
U \int_{ \omega} \int_{\omega^\prime} \int_{ \bar{\omega}} ( \bar{c}_{\omega + \bar{\omega}}
 c_{\omega} ) ( \bar{d}_{\omega^\prime} d_{\omega^{\prime} + \bar{\omega} })
 \nonumber
 \\
 & &
  \hspace{3mm} ( \mbox{frequency transfer $\bar{\omega} = \omega_1^{\prime} - \omega_1$} )
 \label{eq:forward}
 \\
  & = &
- U \int_{ \omega} \int_{\omega^\prime} \int_{ \bar{\omega}} 
 ( \bar{c}_{\omega + \bar{\omega}}  d_{\omega} ) 
 ( \bar{d}_{\omega^\prime} c_{\omega^{\prime} + \bar{\omega} })
 \nonumber
 \\
 & &
  \hspace{3mm} ( \mbox{exchange frequency $\bar{\omega} =  \omega_1 - \omega_2^{\prime}$} )
 \label{eq:exchange}
 \\
 & = & 
   U \int_{ \omega} \int_{\omega^\prime} \int_{ \bar{\omega}} 
 ( \bar{c}_{\omega + \bar{\omega}}  \bar{d}_{-\omega} ) 
 ( {d}_{-\omega^\prime} c_{\omega^{\prime} + \bar{\omega} })
 \nonumber
 \\
 & &
  \hspace{3mm} ( \mbox{total frequency $\bar{\omega} = \omega_1 +  \omega_2$} ).
 \label{eq:Cooper}
 \end{eqnarray}
\end{subequations}
The last three lines can be generated from each other by re-labelling
the frequencies. However, if we impose an ultraviolet cutoff $  | \bar{\omega} | < \Lambda_0 \ll \xi_0$ 
on the bosonic
frequency, each of the above expressions describes a different low-energy scattering process:
forward scattering (\ref{eq:forward}), exchange scattering (\ref{eq:exchange}), and
Cooper scattering (\ref{eq:Cooper}). 
The forward scattering channel can be ignored for our purpose because
the corresponding susceptibility does not exhibit any singularity.
We therefore retain only the exchange and the Cooper channels for small values of the
corresponding energies, which amounts to retaining only the following
low-energy terms in the effective interaction,
 \begin{equation}
 S_{\rm int} \approx \int_{\bar{\omega}} \Theta ( \Lambda_0 - | \bar{\omega} | ) 
 \left[ -U_x \bar{A}_{\bar{\omega}} A_{\bar{\omega} } + U_p  \bar{B}_{\bar{\omega}} B_{\bar{\omega} }
 \right],
 \label{eq:Sint}
 \end{equation}
where we have defined the two composite fields,
 \begin{subequations}
 \begin{eqnarray}
 A_{\bar{\omega}} & = & \int_{\omega} \bar{d}_{\omega} c_{\omega + \bar{\omega}},
 \\
 B_{\bar{\omega}} & = & \int_{\omega} {d}_{-\omega} c_{\omega + \bar{\omega}}.
 \end{eqnarray}
 \end{subequations}
Although $U_x = U_p = U$ we have 
introduced two different coupling constants to distinguish
the bare interaction $U_x$ in the exchange channel from the
interaction $U_p$ in the Cooper (or pairing) channel.
Also note that a certain small sector of 
the three-dimensional frequency space 
formed by $ \omega_1 $, $\omega_2$ and $ \bar{\omega} = \omega_1^{\prime} - \omega_1$
is counted twice in  Eq.~(\ref{eq:Sint}), because
the sectors defined by $ | \omega_2^{\prime} - \omega_1| < \Lambda_0$ and
 $ | \omega_2+  \omega_1| < \Lambda_0$ have a finite overlap
proportional to
 $  \Lambda^2_0 $, corresponding to the intersection of the two
frequency slices shown in Fig.~\ref{fig:slices}.
However, the phase space of these processes
is proportional to $ \Lambda_0^2$ and can  be neglected
if a physical quantity is dominated by generic frequencies.
Finally, we decouple the interaction using two complex bosonic Hubbard-Stratonovich fields
$\chi$ and $\psi$ and obtain for the cutoff-dependent effective low-energy action
 \begin{eqnarray}
 S_{\rm eff} & = & - \int_{\omega}
 \left[ (G_{0}^{c}( i \omega ))^{-1} \bar{c}_{\omega} c_{\omega}
   + (G_{0}^{d}( i \omega ))^{-1} \bar{d}_{\omega} d_{\omega} \right]
  \nonumber
 \\
 & + & \int_{ \bar{\omega}}^{\Lambda_0} \left[ 
 U_x^{-1} \bar{\chi}_{\bar{\omega}} \chi_{\bar{\omega}} +
 \gamma_{x0} ( \bar{A}_{\bar{\omega}} \chi_{\bar{\omega}} + {A}_{\bar{\omega}} \bar{\chi}_{\bar{\omega}} )
 \right]
 \nonumber
 \\
 & + & \int_{ \bar{\omega}}^{\Lambda_0} \left[ U_p^{-1} \bar{\psi}_{\bar{\omega}} \psi_{\bar{\omega}} +
  i \gamma_{p0} ( \bar{B}_{\bar{\omega}} \psi_{\bar{\omega}} + {B}_{\bar{\omega}} \bar{\psi}_{\bar{\omega}} )
 \right],
 \hspace{7mm}
 \label{eq:Seff}
 \end{eqnarray}
where the bare values of the Yukawa vertices are $\gamma_{x0}  = \gamma_{p0} = 1$,
and we have introduced the notation
$ \int_{ \bar{\omega}}^{\Lambda_0} = \int_{\bar{\omega}} \Theta ( \Lambda_0 - | \bar{\omega} | ) $.

\begin{figure}[t]
\includegraphics[width=80mm]{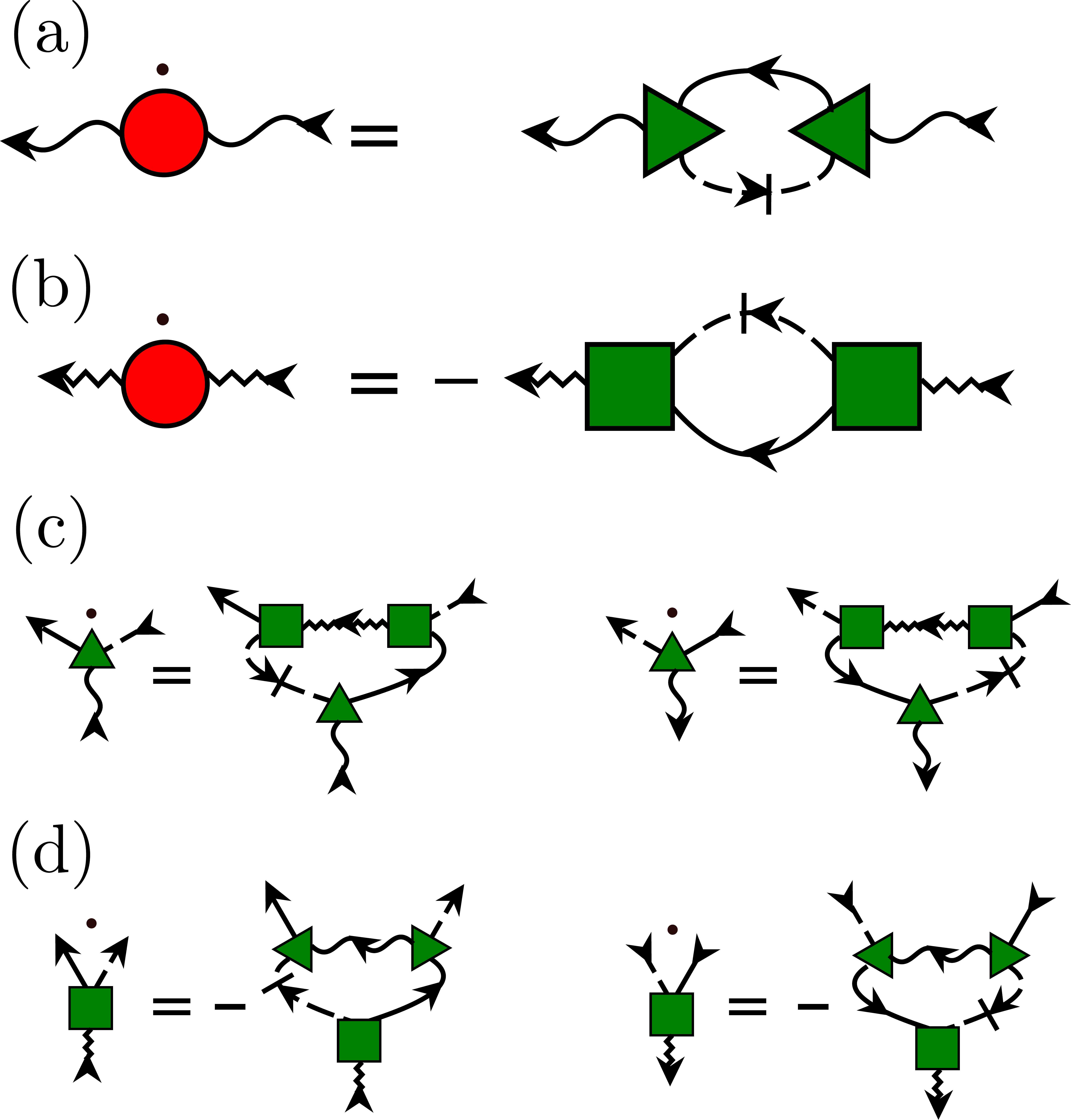}
  \caption{%
(Color online)
Diagram (a) represents  the
FRG flow equation (\ref{eq:Piphflow})
for the self-energy of the Hubbard-Stratonovich field
$\chi$ (wavy line) representing particle-hole fluctuations
in the exchange channel. 
Note that in our cutoff scheme only the 
$d$-electron propagator is regularized.
The shaded triangles and squares denote the 
renormalized Yukawa vertices, while the other symbols are explained
in the caption of Fig.~\ref{fig:flowd}.
Diagram (b) represents the FRG flow equation (\ref{eq:Pippflow}) for the self-energy
of the pairing field $\psi$ (zig-zag line).
In (c) and (d) we show
the FRG flow equations (\ref{eq:vert1}--\ref{eq:vert4}) for the Yukawa vertices associated
with the particle-hole and the particle-particle channel.
}
\label{fig:flowequations}
\end{figure}

\begin{figure}[t]
\includegraphics[width=80mm]{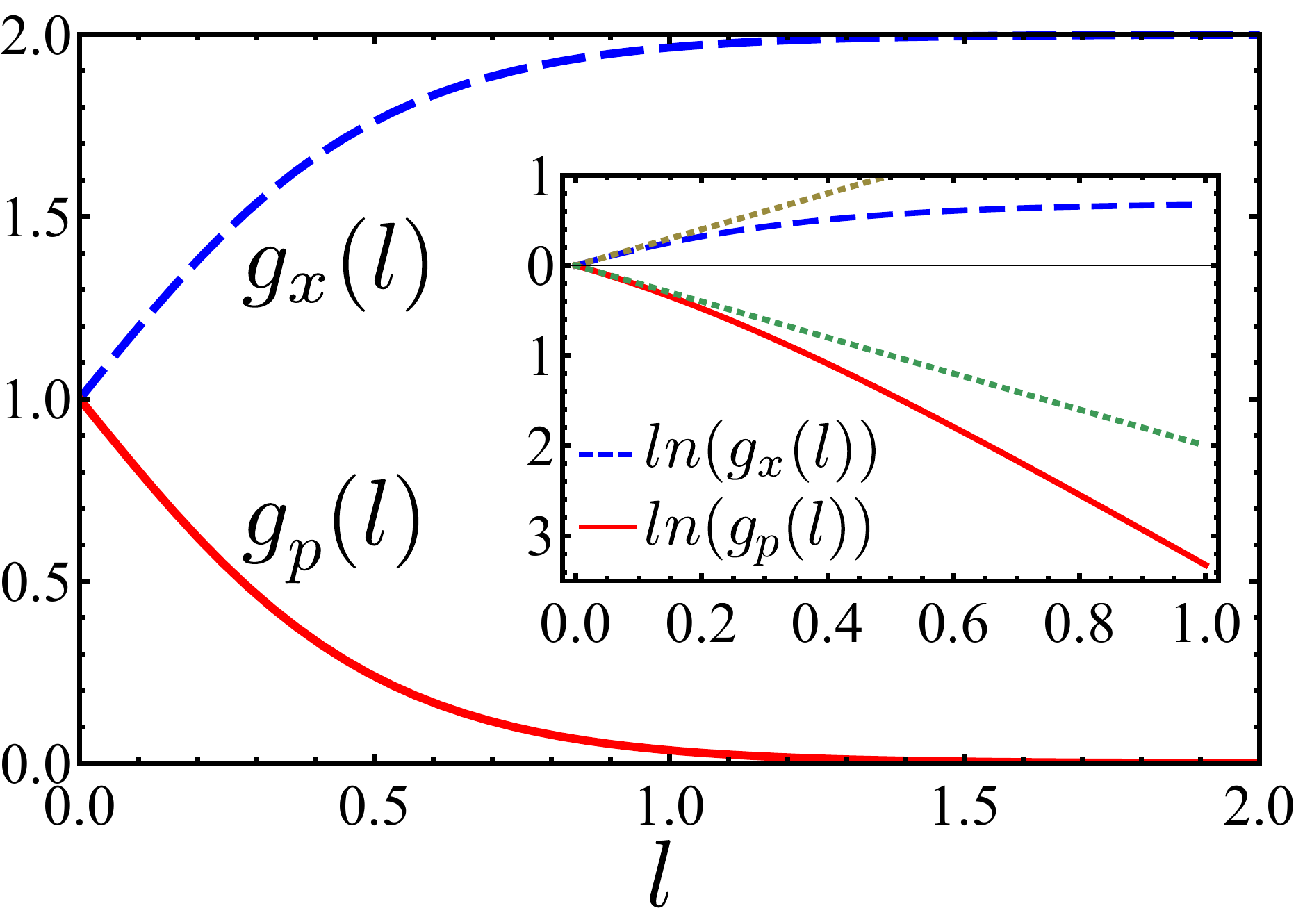}
  \caption{%
(Color online)
Graph of the scale dependent vertex factors $g_x ( l ) = \gamma^2_x ( l )$
and $g_p ( l ) = \gamma^2_p ( l )$ given in (\ref{eq:gxsol}, \ref{eq:gcsol})
as a function of the logarithmic scale factor $l = \ln ( \Lambda_0 / \Lambda )$.
In the inset we show $\ln g_x ( l)$ and $\ln g_p ( l )$  
which in the leading order parquet approximation are replaced by
the straight dashed lines.
}
\label{fig:vertexscale}
\end{figure}

It is now straightforward to write down formally exact FRG
flow equations of the above theory for the irreducible vertices	\cite{Kopietz10,Schuetz05}.
For our purpose, it is sufficient to work with a truncation of these flow equations
where only irreducible vertices with two and three external legs are retained.
Using the same cutoff scheme as in Sec.~\ref{sec:deephole} (i.e., we introduce
a cutoff only into the $d$-electron propagator), the cutoff-dependent propagators $F^{\chi}_{\Lambda}(i \bar{\omega})$ 
and  $F^{\psi}_{\Lambda} ( i \bar{\omega})$ of our bosonic fields are of the form
 \begin{eqnarray}
 F^\chi_{\Lambda} ( i \bar{\omega } ) & = & 
 \frac{1}{ U_x^{-1} + \Pi^{\rm ph}_{\Lambda} ( i \bar{\omega} ) },
 \\
F^\psi_{\Lambda} ( i \bar{\omega } ) & = & 
 \frac{1}{ U_p^{-1} + \Pi^{\rm pp}_{\Lambda} ( i \bar{\omega} ) },
 \end{eqnarray} 
where $\Pi^{\rm ph}_{\Lambda} ( i \bar{\omega} )$ and
 $\Pi^{\rm pp}_{\Lambda} ( i \bar{\omega} )$ are the
cutoff-dependent 
irreducible particle-hole and particle-particle bubbles.
These can be identified with the 
self-energies of our Hubbard-Stratonovich fields. 
From the formally exact FRG flow equation for the generating functional of the
one-line irreducible vertices \cite{Wetterich93,Kopietz10,Metzner12,Schuetz05}
it is now straightforward to write down formally
exact FRG flow equations for the vertices of our coupled fermion-boson theory.
Neglecting the mixed four-point vertices with two bosonic and two fermionic external legs
(these vertices vanish at the initial scale and are expected to remain small at least in the
weak coupling regime) 
the bosonic self-energies satisfy the truncated
FRG flow equations
 \begin{subequations}
 \begin{eqnarray}
 \partial_{\Lambda} \Pi^{\rm ph}_{\Lambda} ( i \bar{\omega} ) & = & 
 \int_{\omega} G^c_{\Lambda} ( i \omega ) \dot{G}^d_{\Lambda} ( i \omega - i
 \bar{\omega} ) 
 \nonumber
 \\
 & \times &
\Gamma^{ \bar{c}d \chi}_{\Lambda} ( \omega , \omega - \bar{\omega} , 
 \bar{\omega } ) 
 \Gamma^{ \bar{d} c \bar{\chi}}_{\Lambda} ( \omega - \bar{\omega} , \omega , \bar{\omega} ),
 \hspace{7mm}
 \label{eq:Piphflow}
 \\
 \partial_{\Lambda} \Pi^{\rm pp}_{\Lambda} ( i \bar{\omega} ) & = & -
 \int_{\omega} G^c_{\Lambda} ( i \omega ) \dot{G}^d_{\Lambda} ( - i \omega + i
 \bar{\omega} ) 
 \nonumber
 \\
 & \times &
\Gamma^{ \bar{c} \bar{d} \psi}_{\Lambda} ( \omega ,  \bar{\omega} - \omega , 
 \bar{\omega } ) 
 \Gamma^{ d c \bar{\psi}}_{\Lambda} ( \bar{\omega} - \omega , \omega , \bar{\omega} ).
 \hspace{7mm}
 \label{eq:Pippflow}
 \end{eqnarray}
 \end{subequations}
A graphical representation of these flow equations is shown
in Fig.~\ref{fig:flowequations} (a) and (b).
The four different Yukawa vertices, appearing in Eqs.~(\ref{eq:Piphflow}) and
(\ref{eq:Pippflow}), satisfy 
the flow equations shown graphically
in  Fig.~\ref{fig:flowequations} (c) and (d). The flow equations for the
Yukawa vertices are explicitly given by
 \begin{subequations}
\begin{widetext}
\begin{eqnarray}
 \partial_{\Lambda}
 \Gamma^{\bar{c} d \chi}_{\Lambda} ( \omega_1^{\prime} , \omega_1 , \bar{\omega}_2 )
 & = &  \int_{\bar{\omega}} F^{\psi}_{\Lambda} ( i \bar{\omega} )  
\dot{G}^d_{\Lambda} ( - i \omega_1^{\prime} + i \bar{\omega} )
G^c_{\Lambda} ( - i  \omega_1 +  i \bar{\omega})
 \nonumber
 \\
 & & \times
 \Gamma^{\bar{c} \bar{d} \psi}_{\Lambda} ( \omega_1^{\prime} , - \omega_1^{\prime} + 
 \bar{\omega} , \bar{\omega} )
  \Gamma^{d c \bar{\psi}}_{\Lambda} ( \omega_1 , - \omega_1 + 
 \bar{\omega} , \bar{\omega} )
  \Gamma^{\bar{c} d  \chi}_{\Lambda} ( - \omega_1  + \bar{\omega}, - \omega_1^{\prime} + 
 \bar{\omega} , \bar{\omega}_2 ) ,
 \label{eq:vert1}
 \\
\partial_{\Lambda}
 \Gamma^{\bar{d} c \bar{\chi}}_{\Lambda} ( \omega_1^{\prime} , \omega_1 , \bar{\omega}_2 )
 & = &  \int_{\bar{\omega}} F^{\psi}_{\Lambda} ( i \bar{\omega} )  
{G}^c_{\Lambda} ( - i \omega_1^{\prime} + i \bar{\omega} )
\dot{G}^d_{\Lambda} ( - i  \omega_1 +  i \bar{\omega})
 \nonumber
 \\
 & & \times
 \Gamma^{\bar{c} \bar{d} \psi}_{\Lambda} (  - \omega_1^{\prime} + 
 \bar{\omega} ,  \omega_1^{\prime} , \bar{\omega} )
  \Gamma^{d c \bar{\psi}}_{\Lambda} ( - \omega_1 + 
 \bar{\omega} , \omega_1 , \bar{\omega} )
  \Gamma^{\bar{d} c  \bar{\chi}}_{\Lambda} ( - \omega_1  + \bar{\omega}, - \omega_1^{\prime} + 
 \bar{\omega} , \bar{\omega}_2 ) ,
 \label{eq:vert2}
 \\
 \partial_{\Lambda}
 \Gamma^{\bar{c} \bar{d} \psi}_{\Lambda} ( \omega_1^{\prime} , \omega_2^{\prime} , 
 \bar{\omega}_1 )
 & = &  - \int_{\bar{\omega}} F^{\chi}_{\Lambda} ( i \bar{\omega} )  
\dot{G}^d_{\Lambda} (  i \omega_1^{\prime} - i \bar{\omega} )
G^c_{\Lambda} (  i  \omega_2^{\prime} +  i \bar{\omega})
 \nonumber
 \\
 & & \times
 \Gamma^{\bar{c} {d} \chi}_{\Lambda} ( \omega_1^{\prime} ,  \omega_1^{\prime} - 
 \bar{\omega} , \bar{\omega} )
  \Gamma^{\bar{d} c \bar{\chi}}_{\Lambda} ( \omega_2^{\prime} ,  \omega_2^{\prime} + 
 \bar{\omega} , \bar{\omega} )
  \Gamma^{\bar{c} \bar{d}  \psi}_{\Lambda} (  \omega_2^{\prime}  + \bar{\omega}, 
 \omega_1^{\prime} - 
 \bar{\omega} , \bar{\omega}_1 ) ,
 \label{eq:vert3}
 \\
 \partial_{\Lambda}
 \Gamma^{d c \bar{\psi}}_{\Lambda} ( \omega_1 , \omega_2 , 
 \bar{\omega}_1 )
 & = &  - \int_{\bar{\omega}} F^{\chi}_{\Lambda} ( i \bar{\omega} )  
{G}^c_{\Lambda} (  i \omega_1 + i \bar{\omega} )
\dot{G}^d_{\Lambda} (  i  \omega_2 -  i \bar{\omega})
 \nonumber
 \\
 & & \times
 \Gamma^{\bar{c} {d} \chi}_{\Lambda} ( \omega_1 + \bar{\omega} ,  \omega_1 
  , \bar{\omega} )
  \Gamma^{\bar{d} c \bar{\chi}}_{\Lambda} ( \omega_2 - \bar{\omega} ,  \omega_2   , \bar{\omega} )
  \Gamma^{d c  \bar{\psi}}_{\Lambda} (  \omega_1  + \bar{\omega}, 
 \omega_2 -  \bar{\omega} , \bar{\omega}_1 ) .
 \label{eq:vert4} 
\end{eqnarray}
\end{widetext}
 \end{subequations}
Note that in our cutoff scheme where only the
$d$-electron propagator is regularized 
the FRG equation for the fermionic self-energy is still
given by Eq.~(\ref{eq:FRG}).
To recover the leading order parquet results for the X-ray response
it is sufficient to
set  all frequency dependencies of the Yukawa vertices equal to zero.
In this limit
 \begin{subequations}
 \begin{eqnarray}
  \Gamma^{\bar{c} d \chi} ( 0, 0, 0 ) & = &  \Gamma^{\bar{d} c \bar{\chi}} ( 0, 0, 0 ) \equiv \gamma_x,
 \\
 \Gamma^{\bar{c} \bar{d} \psi} (0 ,  0, 0 ) 
 & = & \Gamma^{d c \bar{\psi}} (0 ,  0, 0 ) \equiv i \gamma_p.
 \end{eqnarray}
 \end{subequations}
Ignoring all self-energy corrections (fermionic and bosonic) 
to the right-hand sides
of the FRG flow equations we find
 \begin{subequations}
 \begin{eqnarray}
 \partial_l \Pi^{\rm ph} ( 0 ) & = & - \nu_0 \gamma_x^2,
 \label{eq:flowPi}
 \\
  \partial_l \Pi^{\rm pp} ( 0 ) & = &  \nu_0  \gamma_p^2,
 \label{eq:flowPsi}
 \\
  \partial_l \gamma_x & = &  \nu_0 U_p  \gamma_p^2   \gamma_x  ,
 \\
\partial_l \gamma_p & = &  - \nu_0 U_x \gamma_x^2  \gamma_p   ,
 \end{eqnarray}
 \end{subequations}
where $\partial_l = - \Lambda \partial_{\Lambda}$ is the logarithmic scale derivative.
Setting $u_x = \nu_0 U_x$ and $u_p = \nu_0 U_p$ and defining
$g_x =  \gamma_x^2$ and $g_p =  \gamma_p^2$, 
the last two equations can be 
written as
 \begin{subequations}
 \begin{eqnarray}
 \partial_l g_x & = &  2 u_p g_x g_p,
 \label{eq:dif1}
 \\
 \partial_l g_p & = &  - 2 u_x g_x g_p,
 \label{eq:dif2}
 \end{eqnarray}
 \end{subequations}
which should be solved with the initial conditions
$g_x (l=0 ) = g_p (l=0) =1$. 
Using the conservation law $\partial_l [ u_x g_x ( l ) + u_p g_p (l) ] =0$, we can easily solve the two coupled
differential equations (\ref{eq:dif1}) and  (\ref{eq:dif2}) exactly, 
  \begin{subequations}
 \begin{eqnarray}
 g_x (l ) & = & \frac{  (u_x + u_p ) e^{ ( u_x + u_p) l }}{u_x e^{ (u_x + u_p ) l} + u_p e^{
 - ( u_x + u_p )l }},
 \label{eq:gxsolu}
 \\
  g_p ( l ) & = &   \frac{  (u_x + u_p ) e^{ - ( u_x + u_p) l }}{u_x e^{ (u_x + u_p ) l} + u_p e^{
 - ( u_x + u_p )l }}.
 \label{eq:gcsolu}
 \end{eqnarray}
 \end{subequations}
For simplicity, let us now set $ u_x = u_p \equiv u$, so that
the flow of the vertices reduces to
 \begin{subequations}
 \begin{eqnarray}
 g_x (l ) & = & \frac{ e^{ 2 u l }}{\cosh ( 2 u l )},
 \label{eq:gxsol}
 \\
  g_p ( l ) & = & \frac{ e^{ -2 u l }}{\cosh ( 2 u l )}.
 \label{eq:gcsol}
 \end{eqnarray}
 \end{subequations}
A graph of these functions as a function of the logarithmic flow parameter $l$
is shown in Fig.~\ref{fig:vertexscale}.
Substituting our explicit  expressions (\ref{eq:gxsol}, \ref{eq:gcsol})  
for the scale dependent vertices
into Eqs.~(\ref{eq:flowPi}) and (\ref{eq:flowPsi})
we obtain
 \begin{subequations}
 \begin{eqnarray}
 \partial_l \Pi^{\rm pp} ( 0 ) & = & - \nu_0 \frac{ e^{ 2 u l }}{\cosh ( 2 u l )}, 
 \label{eq:flowPi2}
 \\
  \partial_l \Pi^{\rm ph} ( 0 ) & = &  \nu_0 \frac{ e^{ -2 u l }}{\cosh ( 2 u l )}.
 \label{eq:flowPsi2}
 \end{eqnarray}
 \end{subequations}
The crucial point is now that our leading order low-energy 
expansion of the frequency dependent susceptibilities is only valid as long as
$ | \omega | \lesssim \Lambda = \Lambda_0 e^{-l}$.  
To obtain the leading frequency dependence
of the susceptibilities, we should therefore integrate the above equations up to the
logarithmic scale factor $l_{\ast} = \ln ( \Lambda_0 / | \bar{\omega} | )$.
Moreover, assuming  $u l_{\ast} \lesssim 1$ we may
replace the factor $\cosh ( 2 u l )$ in the denominator
of Eqs.~(\ref{eq:flowPi2}) and (\ref{eq:flowPsi2}) by unity.
This corresponds to the leading order parquet approximation discussed
by Roulet, Gavoret, and Nozi\'{e}res \cite{Roulet69}.
In this approximation we obtain
 \begin{eqnarray}
 \Pi^{\rm ph}_{l_{\ast} } ( 0 )  & = & - \nu_0 \int_0^{l_{\ast}} d l e^{ 2 u l } = 
   \frac{ \nu_0}{2u} [ 1 - e^{ 2 u l^{\ast}}  ]
 \nonumber
 \\
 & = & 
   \frac{ \nu_0}{2u} \left[  1 - \left( \frac{\Lambda_0}{| \bar{\omega} | } \right)^{2u}  \right],
 \\
\Pi^{\rm pp}_{l_{\ast} } ( 0 )  & = &  \nu_0 \int_0^{l_{\ast}} d l e^{ -2 u l } = 
   \frac{ \nu_0}{2u} [ 1- e^{ -2 u l^{\ast}}  ]
 \nonumber
 \\
 & = & 
   \frac{ \nu_0}{2u} \left[  1 - \left( \frac{| \bar{\omega} | }{\Lambda_0} \right)^{2u}  \right].
 \end{eqnarray}
 Finally, we perform an analytic continuation to the real frequency axis and obtain
for the singular part of the retarded particle-hole response function
for positive frequencies,
 \begin{eqnarray} 
 \chi^{\rm ph} ( \omega ) & = & - {\rm Re}  \left[ \left. \Pi^{\rm ph}_{l_{\ast} } ( 0 ) 
 \right|_{ i \bar{\omega} \rightarrow \omega + i 0^+ } \right]  
 \nonumber
 \\
 & = & 
\frac{ \nu_0}{2u} \left[  \left( \frac{\Lambda_0}{\omega } \right)^{2u} -1 \right].
 \end{eqnarray}
Identifying $\Lambda_0 = \xi_0$ and
keeping in mind that we have set the  threshold frequencies
equal to zero, we reproduce  
the result (\ref{eq:responsealpha})  which has first been obtained by
Mahan \cite{Mahan67} and later by
Roulet, Gavoret, and Nozi\'{e}res \cite{Roulet69} using conventional
parquet methods.
Within our FRG formalism, it is straightforward
to calculate also the leading singularity of the 
particle-particle response function; we obtain
  \begin{eqnarray} 
 \chi^{\rm{pp}} ( \omega ) & = &   {\rm Re}  \left[ \left. \Pi^{\rm pp}_{l_{\ast} } ( 0 ) 
 \right|_{ i \bar{\omega} \rightarrow \omega + i 0^+ } \right]
 \nonumber
 \\
  & = &
\frac{ \nu_0}{2u} \left[  1 - \left( \frac{\omega}{\Lambda_0 } \right)^{2u}  \right].
 \label{eq:chipp}
 \end{eqnarray}
Note that the leading frequency dependence vanishes in a  non-analytic 
way for $\omega \rightarrow 0$. We have not been able to find
Eq.~(\ref{eq:chipp}) anywhere in the published literature on the X-ray problem.

\section{\label{sec:conclusions} Conclusions}

In this work we have shown how to obtain the threshold behavior
of the deep hole Green function as well as the particle-hole and 
particle-particle response functions in the X-ray problem
using the functional renormalization group.
While our FRG results  are equivalent to the leading order
parquet approximation \cite{Roulet69},  the calculational effort  
within the framework of the FRG is much lower than
in the traditional parquet approach.
Technically, a novel feature of our approach is the use of multi-component
Hubbard-Stratonovich fields to take into account the singularities
in competing scattering channels. 
Note that 
recently several authors have proposed
parametrizations of the
momentum- and frequency-dependent  four-point vertex
in a purely fermionic formulation of the FRG
which take the special combinations of momenta and frequencies
associated with bosonic collective 
modes \cite{Hedden04,Karrasch08,Husemann09,Jakobs10,Husemann12,Giering12,Bauer14} into account.
The method based on  multi-channel
Hubbard-Stratonovich fields proposed here is an alternative to these purely
fermionic implementations of the FRG. 
In fact, 
our approach is similar in spirit to the version of the
parquet renormalization group discussed in Ref.~[\onlinecite{Maiti13}],
which is also based on the explicit introduction of fermion-boson vertices via
Hubbard-Stratonovich transformations.
We believe that our method will also be useful
in the context of other problems where
many-body interactions lead to
competing instabilities in  more than one channel. 
For example, in two-dimensional Fermi systems
the low-energy scattering processes 
of interacting fermions can be classified into
forward-, exchange- and Cooper scattering \cite{Shankar94,Belitz97,Drukier15},
which can be bosonized by introducing
three different Hubbard-Stratonovich fields. 
It is straightforward to write down the coupled FRG flow equations for the
irreducible two-point and three-point vertices for this problem.
The corresponding
FRG flow will be discussed elsewhere.
Finally, let us point out that with the FRG it is straightforward
to go beyond the leading
order parquet approximation by taking self-energy corrections to the
propagators and higher order irreducible vertices into account.

\section*{ACKNOWLEDGMENT}
Two of us (P. L. and P. K.) gratefully acknowledge
financial support by the DFG via
FOR 723.


\begin{thebibliography}{99}
%
%
\bibitem{Fetter71}
See, for example, A. L. Fetter and J. D. Walecka, 
{\it{Quantum Theory of Many-Particle Systems}}, (McGraw-Hill, New York, 1971).
%
% \bibitem{deDominicis64}
% C. de Dominicis and P. C. Martin, J. Math. Phys. {\bf{5}}, 14 (1964).
%
\bibitem{Sudakov56}
V. V. Sudakov, Dokl. Akad. Nauk SSSR {\bf{111}}, 338 (1956)
[Sov. Phys. Doklady {\bf{1}}, 662 (1956)].
%
\bibitem{Diatlov57}
I. T. Diatlov, V. V. Sudakov, and K. A. Ter-Martirosian,
Zh. Eksp. Theor. Fiz. {\bf{32}}, 767 (1957)
[Sov. Phys. JETP {\bf{5}}, 631 (1957)].
%
\bibitem{Abrikosov65}
A. A. Abrikosov, Physics {\bf{2}}, 5 (1965).
%
\bibitem{Roulet69}
B. Roulet, J. Gavoret, and P. Nozi\`{e}res,
Phys. Rev. {\bf{178}}, 1072 (1969).
%
\bibitem{Nozieres69}
P. Nozi\`{e}res, J. Gavoret, and B. Roulet, Phys. Rev. {\bf{178}}, 1084 (1969).
%
\bibitem{Fukushima71}
K. Fukushima, Prog. Theor. Phys. {\bf{46}}, 1307 (1971).
%
\bibitem{kleinert95}
P. Kleinert and H. Schlegel, Physica A {\bf{218}}, 507 (1995).
%
\bibitem{Janis99}
V. Jani\v{s}, Phys. Rev. B {\bf{60}}, 11345 (1999);
V. Jani\v{s} and P. Augustinsky, Phys. Rev. B {\bf{75}}, 165108 (2007). 
%
\bibitem{Dzyaloshinskii72}
I. E. Dzyaloshinskii and A. I. Larkin, 
Zh. Eksp. Teor. Fiz. {\bf{61}}, 791 (1971)
[Sov. Phys. JETP {\bf{34}}, 422 (1972)].
%
\bibitem{Gorkov74}
L. P. Gorkov and I. E. Dzyaloshinskii, Zh. Eksp. Teor. Fiz. {\bf{67}}, 397 (1974)
[Sov. Phys. JETP {\bf{40}}, 198 (1975)].
%
\bibitem{Dzyaloshinskii87}
 I. E. Dzyaloshinskii, Zh. Eksp. Teor. Fiz. {\bf{93}}, 1487 (1987)
[Sov. Phys. JETP {\bf{66}}, 848 (1987)].
%
\bibitem{Zheleznyak97}
A. T. Zheleznyak, V. M. Yakovenko, and I. E. Dzyaloshinskii,
Phys. Rev. B {\bf{55}}, 3200 (1997).
%
\bibitem{Babu73}
S. Babu and G. E. Brown, Ann. Phys. (N. Y.) {\bf{78}}, 1 (1973).
%
\bibitem{Jackson82}
A. D. Jackson, A. Lande, and R. A. Smith, Phys. Rep. {\bf{86}}, 55 (1982).
%
\bibitem{Quader87}
K. Quader, K. Bedell, and G. E. Brown, Phys. Rev. B {\bf{36}}, 156 (1987). 
%
\bibitem{Pfitzner87}
M. Pfitzner and P. W\"{o}lfle, Phys. Rev. B {\bf{35}}, 4699 (1987).
%
\bibitem{Yeo96}
J. Yeo and M. A. Moore, Phys. Rev. B {\bf{54}}, 4218 (1996). 
%
\bibitem{Yasuda99} 
K. Yasuda, Phys. Rev. A {\bf{59}}, 4133 (1999).
%
\bibitem{Bergli10}
E. Bergli and M. Hjorth-Jensen, Ann. Phys. (N.Y.) {\bf{326}}, 1125 (2011).
%
% \bibitem{Bickers89}
% N. E. Bickers and D. J. Scalapino, Ann. Phys. {\bf{193}}, 206 (1989).
%
\bibitem{Bickers91}
N. E. Bickers and S. R. White, Phys. Rev. B {\bf{43}}, 8044 (1991).
%
\bibitem{Bickers92}
N. E. Bickers and D. J. Scalapino, Phys. Rev. B {\bf{46}}, 8050 (1992).
%
\bibitem{Irkhin01}
V. Yu. Irkhin, A. A. Katanin, and M. I. Katsnelson,
Phys. Rev. B {\bf{64}}, 165107 (2001).
%
\bibitem{Chubukov08}
A. V. Chubukov, D. V. Efremov, and I. Eremin,
Phys. Rev. B {\bf{78}}, 134512 (2008).
%
\bibitem{Maiti10}
S. Maiti and A. V. Chubukov, Phys. Rev. B {\bf{82}}, 214515 (2010).
%
\bibitem{Maiti13}
S. Maiti and A. V. Chubukov, arXiv:1305.4609, to be published in
{\it{Proceedings of the XVII Training Course in the physics of Strongly Correlated 
Systems (Vietri sul Mare (Salerno), Italy).}}
%
\bibitem{Nandkishore12}
R. Nandkishore, R. Thomale, and A. V. Chubukov, Phys. Rev. B {\bf{89}}, 144501 (2014).
%
\bibitem{Tam13}
K.-M. Tam, H. Fotso, S.-X. Yang, T.-W. Lee, J. Moreno, J. Ramanujam, and M. Jarrell, Phys. Rev. E {\bf{87}}, 013311 (2013).
%
\bibitem{Mahan67}
G. D. Mahan, Phys. Rev. {\bf{163}}, 612 (1967); see also
G. D. Mahan, {\it{Many-Particle Physics}}, (Kluwer Academic/Plenum Publishers,
New York, 2010, 3rd edition).
%
\bibitem{Nozieres69b}
P. Nozi\`{e}res, and C. T. De Dominicis, Phys. Rev. B {\bf{178}}, 1097 (1969).
%
\bibitem{Gogolin99}
A. O. Gogolin, A. A. Nersesyan, and A. M. Tsvelik, {\it{Bosonization and Strongly Correlated Systems} (Cambridge University Press, Cambridge, England, 1999).}
%
\bibitem{Smith88}
R. A. Smith and A. Lande, in {\it{Condensed Matter Theories}}, Vol. 3, edited by J. S. Arponen, R. F. Bishop, and M. Manninen (Plenum, New York, 1988).
%
\bibitem{Kopietz01}
P. Kopietz and T. Busche, Phys. Rev. B {\bf{64}}, 155101 (2001).
%
\bibitem{Salmhofer01}
M. Salmhofer and C. Honerkamp, Prog. Theor. Phys. {\bf{105}}, 1 (2001).
%
\bibitem{Kopietz10}
P. Kopietz, L. Bartosch, and F. Sch\"{u}tz, {\it{Introduction to the Functional Renormalization
Group}}, (Springer, Berlin, 2010).
%
\bibitem{Metzner12}
W. Metzner, M. Salmhofer, C. Honerkamp, V. Meden, and
K. Sch\"{o}nhammer, Rev. Mod. Phys. {\bf{84}}, 299 (2012).
%
\bibitem{Shifman99}
For a single Grassman variable $\eta$ the Dirac delta-function is defined by $\delta ( \eta ) = \eta$.
This 
has all the properties of a delta-function, as discussed, for
example, by M. A. Shifman, {\it{ITEP Lectures
on Particle Physics and Field Theory}},  Volume I, (World Scientific, Singapore, 1999).
The product $\prod_{\tau}$ in Eq.~(\ref{eq:deltagras}) should be understood
as a properly regularized product over discretized time steps.
%
\bibitem{Litim01}
D. Litim, Phys. Rev. D {\bf{64}}, 105007 (2001).
% 
\bibitem{Schuetz05}
F. Sch\"{u}tz, L. Bartosch, and P. Kopietz, Phys. Rev. B {\bf{72}}, 035107 (2005).
%
\bibitem{Wetterich93}
C. Wetterich, Phys. Lett. B {\bf{301}}, 90 (1993).
%
\bibitem{Hedden04}
R. Hedden, V. Meden, T. Pruschke, and K. Sch\"{o}nhammer, J. Phys.: Condens. Matter {\bf{16}}, 5279 (2004).
%
\bibitem{Karrasch08}
C. Karrasch, R. Hedden, R. Peters, T. Pruschke, K. Sch\"{o}nhammer, and V. Meden, J. Phys.: Condens. Matter {\bf{20}}, 345205 (2008).
%
\bibitem{Husemann09}
C. Husemann and M. Salmhofer, Phys. Rev. B {\bf{79}}, 195125 (2009).
%
\bibitem{Jakobs10}
S. Jakobs, M. Pletyukhov, and H. Schoeller,
Phys. Rev. B {\bf {81}}, 195109 (2010).
%
\bibitem{Husemann12}
C. Husemann, K.-U. Giering, and M. Salmhofer, Phys. Rev. B {\bf{85}}, 075121 (2012).
%
\bibitem{Giering12}
K.-U. Giering and M. Salmhofer, Phys. Rev. B {\bf{86}}, 245122 (2012).
%
\bibitem{Bauer14}
F. Bauer, J. Heyder, and J. von Delft, Phys. Rev. B {\bf{89}}, 045128 (2014).
%
\bibitem{Shankar94}
R. Shankar, Rev. Mod. Phys. {\bf{66}}, 129 (1994).
%
\bibitem{Belitz97}
D. Belitz, T. R. Kirkpatrick, and T. Vojta, Phys. Rev. B {\bf{55}}, 9452 (1997).
%
\bibitem{Drukier15}
C. Drukier, P. Lange, and P. Kopietz, Eur. Phys. J. B {\bf{88}}, 41 (2015).
%

\end{thebibliography}
\end{document}